\documentclass{astlb}
\usepackage{graphicx}
\usepackage{natbib}
\usepackage{color}

\usepackage[hyperfootnotes=false]{hyperref}

\usepackage{multirow}

\definecolor{darkblue}{rgb}{0,0,0.9}

\def\smfigure#1#2#3{
  \begin{minipage}{1.0\columnwidth}
    \begin{minipage}{0.049\columnwidth}
      \rotatebox{90}{\small\phantom{0000}#3}
    \end{minipage}
    \begin{minipage}{0.9\columnwidth}
      \includegraphics[viewport=40 188 556 678,width=0.97\columnwidth]{#1}
      \centerline{\small #2}
    \end{minipage}

    \vskip 3pt
    ~
  \end{minipage}
}

\def\plotspectrum#1#2#3{
  \begin{minipage}{1.0\columnwidth}
    \begin{minipage}{0.049\columnwidth}
      \rotatebox{90}{\phantom{0000}#3}
    \end{minipage}
    \begin{minipage}{0.95\columnwidth}
      \includegraphics[viewport=58 188 556 497,width=0.97\columnwidth]{#1}
      \centerline{#2}
    \end{minipage}
    
    \vskip 3pt
    ~
  \end{minipage}
}

\begin{document}

\journalinfo{2016}{42}{4}{240}[250]

\title{Sample of Cataclysmic Variables from 400d X-ray Survey}

\author{R.~A.~Burenin\email{rodion@hea.iki.rssi.ru}\address{1},
  M.~G.~Revnivtsev\address{1}, A. Yu. Tkachenko\address{1},
  V.~S.~Vorobyev\address{1}, A.~N.~Semena\address{1}, A.~V.~Meshcheryakov\address{1},
  S.~N.~Dodonov\address{2}, M.~V.~Eselevitch\address{3},
  M.~N.~Pavlinsky\address{1}
  \addresstext{1}{Space Research Institute RAS (IKI), Moscow, Russia}
  \addresstext{2}{Special Astrophysical Observatory RAS, Nizhnij Arkhyz, Russia}
  \addresstext{3}{Institute of Solar-Terrestrial Physics SB RAS, Irkutsk, Russia}}

\shortauthor{R. A. Burenin et al.}

\shorttitle{CV sample from 400d X-ray survey}

\submitted{November 24, 2015}

\begin{abstract}  
  We present a sample of cataclysmic variables (CVs) identified among
  the X-ray sources from the 400 square degree X-ray survey based on
  ROSAT pointing data (400d). The procedure of the CV selection among
  the X-ray sources using additional optical and infrared data from
  Sloan Digital Sky Survey and WISE survey is described. The results
  of the optical observations of the selected objects carried out
  mainly with the Russian--Turkish 1.5-m telescope (RTT-150) and the
  6-m telescope of the Special Astrophysical Observatory of the
  Russian Academy of Sciences (BTA) are presented.  Some observations
  have also been performed with the Sayan Observatory 1.6-m AZT-33IK
  telescope. Currently we selected eight CVs, four of which were found
  for the first time in our work. Based on this sample, we have
  obtained preliminary constraints on the CV X-ray luminosity function
  in the solar neighborhood in the low luminosity range,
  $L_X\sim 10^{29}$--$10^{30}$~erg~s$^{-1}$ (0.5--2~keV). We show that
  the logarithmic slope of the CV X-ray luminosity function in this
  luminosity range is less steep than at
  $L_X>10^{31}$~erg~s$^{-1}$. From our CV X-ray luminosity function
  estimates it follows that few thousand CVs will be detected in the
  Spectrum-R\"ontgen-Gamma (SRG) observatory all-sky X-ray survey at
  high Galactic latitudes, which will allow to obtain much more
  accurate measurements of CV X-ray luminosity function in the
  luminosity range $L_X < 10^{30}$--$10^{31}$~erg\,s$^{-1}$.
  

\end{abstract}

\section{Introduction}

Studies of samples of binary systems with accreting white dwarfs
(cataclysmic variables, CVs) provide important information on physical
processes in these systems that are difficult or even impossible to
investigate by other means. For example, it turns out that the main
mechanism of angular momentum losses for CVs with periods shorter than
2--3~h is the emission of gravitational waves
\citep{faulkner71,paczynski81}. This prediction is confirmed by
characteristic features in the properties of the CV population, such
as the minimum in the distribution of orbital periods,
$P_{\mathrm min}\approx 80$~min \citep{gansicke09}, which agree well
with the improved theoretical estimates of this quantity for the case
of angular momentum losses trough gravitational radiation in binary
system \citep{knigge11}.

Nevertheless, there are also some discrepancies between observations
and theoretical views. For example, theoretical modeling of the CV
population shows that most of them must be in a state with a low
accretion luminosity, and their companions must be degenerate objects
\citep[the so-called ``post-bounce'' CVs, e.g.,][]{kolb93,howell01,knigge11}.
However, only a small number of possible post-bounce CV candidates are
known to date \citep[see, e.g.,][]{littlefair08,aviles10}. Among the
problems that can be solved using studies of the statistically
complete samples of CVs, there is a question of how and at what rate
the masses of the white dwarfs (WDs) in these systems grow
\citep[e.g.,][]{zorotovic11}. Since accreting white dwarfs are
apparently the progenitors of type Ia supernovae, all such questions
appear to be closely related to the studies of the properties of type
Ia supernovae, which are used as ``standard candles'' for cosmological
measurements.

It is also turns out that CVs contribute significantly to the Galactic
ridge X-ray emission \citep{revnivtsev09} and to the total X-ray
luminosity of the Galaxy and other galaxies after the subtraction of
the contribution from low-mass X-ray binaries
\citep{sazonov06,revnivtsev08a}. To obtain more accurate measurements
of the contribution from the CV population to the total X-ray
luminosity of galaxies, the studies of the statistically complete
samples of these objects and the data that allow to measure the X-ray
luminosity function are also required.

The selection of the statistically complete CV samples in X-rays is
one of the most natural approaches to obtain unbiased CV samples,
because the X-ray emission is generated in the accretion flow near the
WD surface and, therefore, is a property of all types of CVs (for a
more detailed discussion, see below). Such samples can be obtained in
various X-ray sky surveys. The X-ray luminosity functions of CVs were
measured earlier using the data from the RXTE all-sky survey
\citep{sazonov06} and the ROSAT North Ecliptic Pole survey
\citep{pretorius07b,pretorius12} as well as in harder X-ray band,
using INTEGRAL \citep{revnivtsev08b} and Swift \citep{pretorius14}
data.

Almost all the objects detected in these surveys have X-ray
luminosities\footnote{Here and below all X-ray fluxes and luminosities
  are given in 0.5--2 keV energy band.}
$L_X>10^{30}$~erg~s$^{-1}$. Since the CV X-ray luminosity function
increases toward lower luminosities, the measurements of the CV number
density at luminosities $L_X<10^{30}$~erg~s$^{-1}$ should allow to
refine the contribution from the CV population to the total X-ray
luminosity of the Galaxy. In addition, at those low luminosities, one
could expect to detect the large number of post-bounce CVs.

Thus, the CV X-ray luminosity function measurements at low X-ray
luminosities $L_X<10^{30}$~erg~s$^{-1}$ are of great interest. To
obtain these measurements one should use deeper X-ray surveys, with
flux limits $\sim 10^{-15}$--$10^{-14}$~erg\,s$^{-1}$cm$^{-2}$.  In
this work for these purposes we used the 400 square degree (400d)
X-ray survey, based on ROSAT pointing data, which were used earlier to
detect galaxy clusters \citep[][]{400d}. More than 37\,000 point X-ray
sources with fluxes above $10^{-14}$~erg\,s$^{-1}$cm$^{-2}$ were
detected in this survey; about 22\,000 of these sources were detected
in the ROSAT fields overlapped with Sloan Digital Sky Survey
\citep{400dpointsrc16}. The optical observations with the aim to
search the CVs among the X-ray sources from this survey were started
by our group earlier \citep{ayut15}; in this paper, we discuss the
sample of CVs obtained to date.

Below, we describe the procedure of the CV selection among the ROSAT
X-ray sources using additional optical and infrared data from SDSS and
WISE surveys. We provide a list of CVs selected in the 400d survey to
date and discuss the properties of these objects. We also discuss the
preliminary constraints on the X-ray luminosity function of CVs in the
solar neighborhood obtained using this sample.

\section{CV Selection in X-rays and optical}

X-ray surveys provide one of the most efficient methods to select CVs.
Indeed, all types of CVs appear to be the sources of X-ray emission
associated with the optically thin plasma emission of the accretion
flow heated near the WD surface \citep[see, e.g.,][]{aizu73}.
Depending on the specific physical picture of accretion, the fraction
of the X-ray emission at energies above 0.5~keV in the bolometric
luminosity of CVs can differ significantly, but it always remains
fairly large, at least $\sim1$\% \citep[see, e.g.,][]{BeuermannThomas93}.

\begin{figure}
  \centering
  \includegraphics[width=\linewidth,viewport=22 170 583 705]{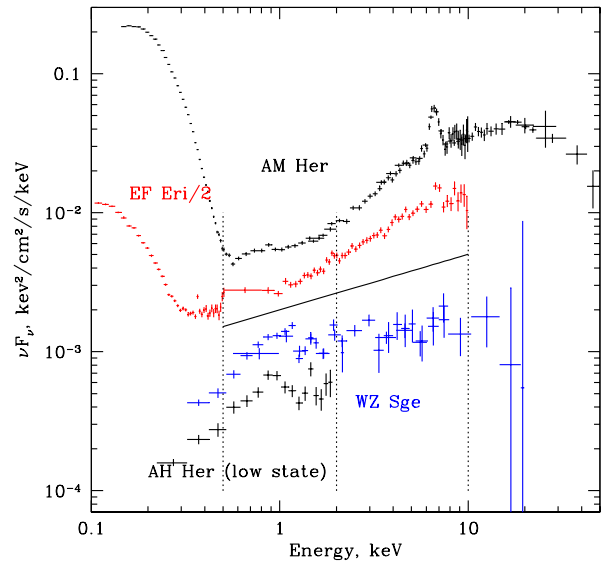}
  \caption{Examples of the CV X-ray spectra. The solid line indicates
    a power-law spectrum $dN/dE\propto E^{-1.6}$.}
  \label{fig:cv_spec}
\end{figure}

The spectrum of CVs at the energies above 0.5~keV is dominated mainly
by the optically thin plasma emission that originates either in an
accretion column or in an optically thin boundary layer near the WD
surface. The hot region near the WD surface is significantly
inhomogeneous in temperature, but the bulk of the X-ray emission is
generated at temperatures of 1--20~keV, producing a spectrum with a
characteristic power-law slope $dN/dE\propto E^{-1.6}$. It is shown in
Fig.~\ref{fig:cv_spec}, where the examples of CV X-ray spectra from
ROSAT, ASCA, and RXTE data are presented.

Fig.~\ref{fig:cv_spec} shows the spectra of the magnetized WDs EF\,Eri
and AM\,Her (the latter in two states: in the high state, in which a
powerful soft component with a temperature of about 60~eV is seen, and
in the low state, when there is no such component in the spectrum) as
well as the spectrum of WZ\,Sge, an accreting WD without a strong
magnetic field. A more detailed discussion of the X-ray spectra for
accreting WDs with a weak magnetic field can be found, for example, in
\cite{byckling10}. It is clearly seen that the CV spectra in the
0.5--10~keV energy band look approximately similar in shape,
irrespective of the CV type and state, and that the general form of
these spectra can be approximately described by a power law within a
photon index of $-1.6$.

\subsection{X-ray data}

In our work we used the ROSAT pointed data in the 0.5--2~keV energy
band, at high galactic latitudes, which were used earlier to search
for distant galaxy clusters \citep[400d survey,][]{400d}. In 400d
survey point X-ray sources were detected only in the central part of
the ROSAT field of view at distances $<18.5\arcmin$ from its center,
where ROSAT angular resolution is better than $70\arcsec$ (PSF
FWHM). The positional accuracy for an X-ray source is, on average, is
better than $10\arcsec$ (at 95\% confidence). We used 1605 ROSAT
pointings at high Galactic latitudes $|b|>25^\circ$. The CVs nearest
to the Sun located within the Galactic disk must be observed at such
latitudes.

The geometric area of the 400d survey for point sources is 436.7 sq.\
deg. The geometric area of the 400d survey overlapping with SDSS
photometric fields is 262.3 sq.\ deg. Since the exposure distribution
of ROSAT pointings is wide, the flux limit turns out to be different
in different fields of the survey. Half and 5\% of the geometric area
are gathered at X-ray flux of $2.5\cdot10^{-14}$ and
$8\cdot10^{-15}$~erg\,s$^{-1}$\,cm$^{-2}$, respectively
\citep{400dpointsrc16}.

\subsection{Optical and Infrared data}

As it was discussed above, in total more than 37\,000 X-ray sources
were detected in the 400d survey, about 22 000 of them --- in the
fields where there is an overlap with SDSS photometric
fields. Obviously, the complete optical identification of all X-ray
sources from the 400d survey is too difficult, and additional data are
required to select CVs. For these purposes, we used the data from the
12-th release of the SDSS \citep{sdss12} and the WISE infrared all-sky
survey \citep{wright10}.

The stellar companions of the WDs in CVs are mostly low-mass
($<\!\! 1M_\odot$) main-sequence stars \citep{knigge06}. However, the WD in
a CV and the accretion disk around it have a temperature no lower than
$10^4$\,K \citep[see, e.g.][]{townsley09}, and, hence, are blue in the
optical band. In the blue part of the optical spectrum and in the near
ultraviolet, the CV spectrum must be always dominated by the emission
from the WD and the accretion disk; therefore, this property can be
used to select CVs in the optical band \citep[see,
e.g.,][]{green82,szkody02}. For the selection of CV candidates, we
used the criterion $u^\prime-g^\prime<0.7$, which was used previously
among other criteria in the SDSS to eliminate the WDs from the
spectroscopic sample of quasars (Richards et al. 2002).

To eliminate a large number of quasars, which strongly contaminate the
CV sample, we additionally discarded the objects with colors
$w_1-w_2>0.6$, where $w_1$ and $w_2$ are the photometric bands of the
WISE infrared all-sky survey with central wavelengths of 3.4 and
4.6~$\mu$m \citep{wright10}. For the vast majority of stars (except
for the cool T dwarfs) and, consequently, for the vast majority of
CVs, this spectral range corresponds to the Rayleigh-Jeans part of the
spectrum; therefore, the CV color must be $w_1-w_2\approx 0$. The
criterion we used is fairly conservative; it should allow the possible
contribution of the companion star to be taken into account even if it
is an L-type dwarf \citep[see, e.g.,][]{schmidt15}.

For the subsequent additional optical observations, we selected CV
candidates with $g^\prime$ magnitudes $<20.0$. Even when the
contribution of the accretion disk is small and the WD contribution
dominates in the spectra of CVs, their absolute magnitudes are
$M_{g^\prime}\approx 12$ \citep{gansicke09,mikej14}. Therefore, the
magnitude $g^\prime\approx20.0$ for such systems corresponds to a
distance of $\approx 400$~pc, which is approximately equal to the
thickness of the Galactic disk. The minimum absolute magnitude of CVs
depends strongly on the WD mass and can reach $M_{g^\prime}\approx 14$
for a WD with a mass of $\approx 1.2 M_\odot$ \citep{mikej14}. Even
for such systems the magnitude limit $g^\prime<20.0$ corresponds to a
distance of $\sim 160$~pc, which also provides an appreciable search
volume.

There are 53 objects satisfying the criteria discussed above in the
400d survey. Some of them (four objects) turned out to be previously
known CVs. To identify the nature of the remaining objects, we carried
out additional optical observations.

\begin{figure*}
  \centering
  \begin{tabular}[c]{ll}
    \phantom{00000}400d~j$001912.9\!+\!220736$ &\phantom{00000}400d~j$152212.8\!+\!080338$ \\
    \plotspectrum{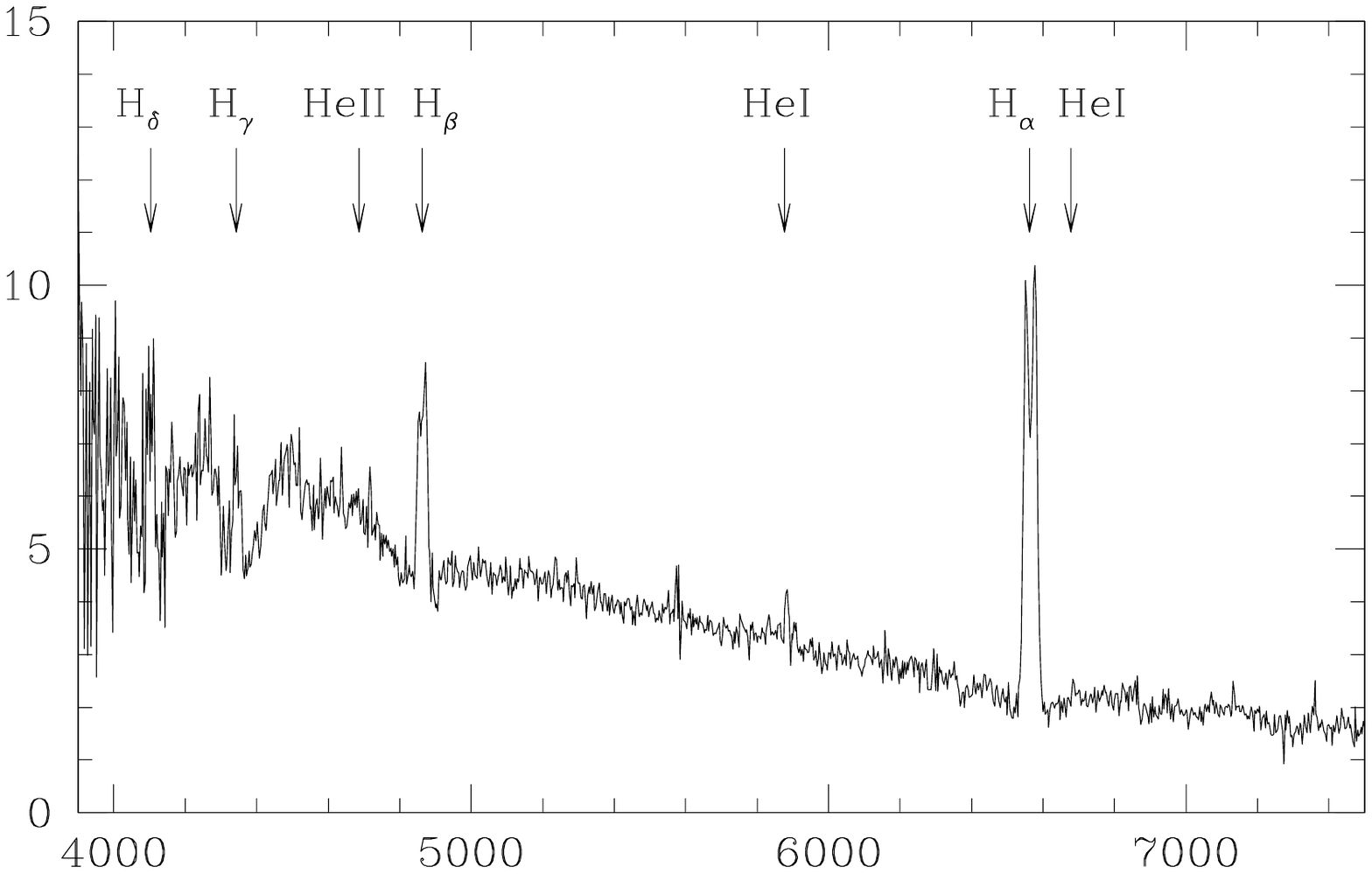}{$\lambda$, \AA}{$F_\lambda, \times
      10^{-17}$~erg~s$^{-1}$~cm$^{-2}$~\AA$^{-1}$} & 
    \plotspectrum{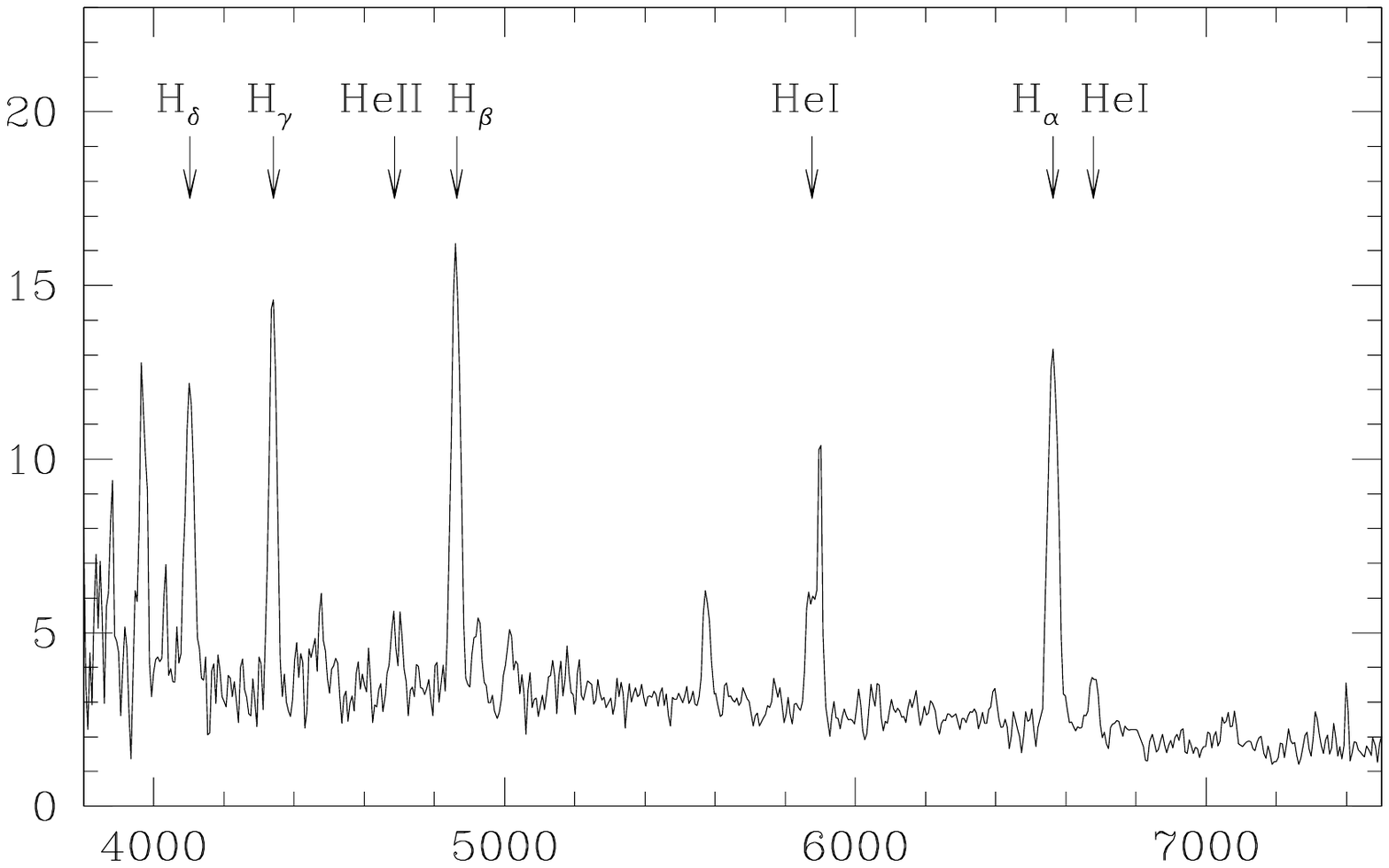}{$\lambda$, \AA}{$F_\lambda, \times
      10^{-17}$~erg~s$^{-1}$~cm$^{-2}$~\AA$^{-1}$} \\
    \phantom{00000}400d~j$154730.1\!+\!071151$ &\phantom{00000}400d~j$160547.5\!+\!240524$ \\
    \plotspectrum{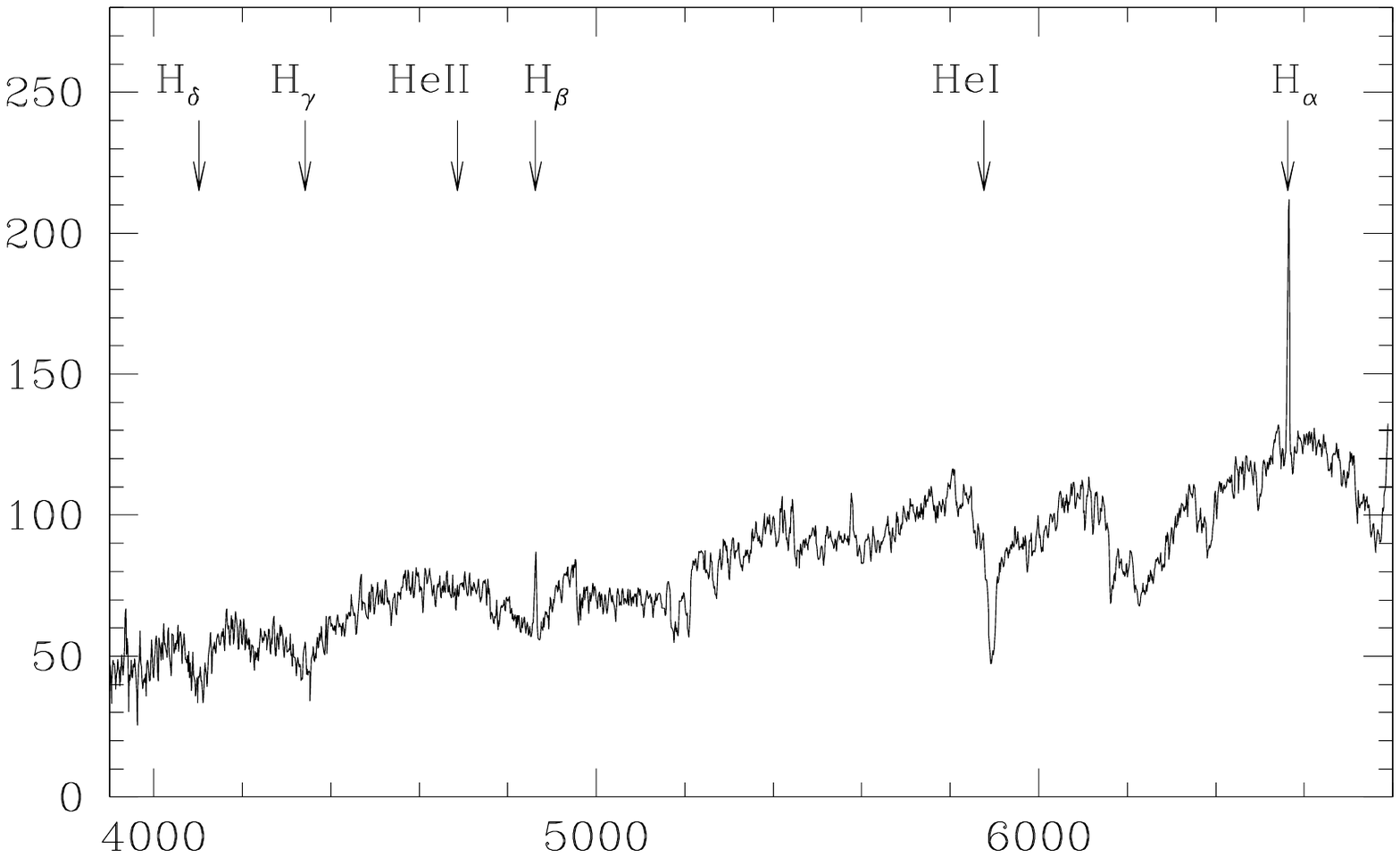}{$\lambda$, \AA}{$F_\lambda, \times
      10^{-17}$~erg~s$^{-1}$~cm$^{-2}$~\AA$^{-1}$} & 
    \plotspectrum{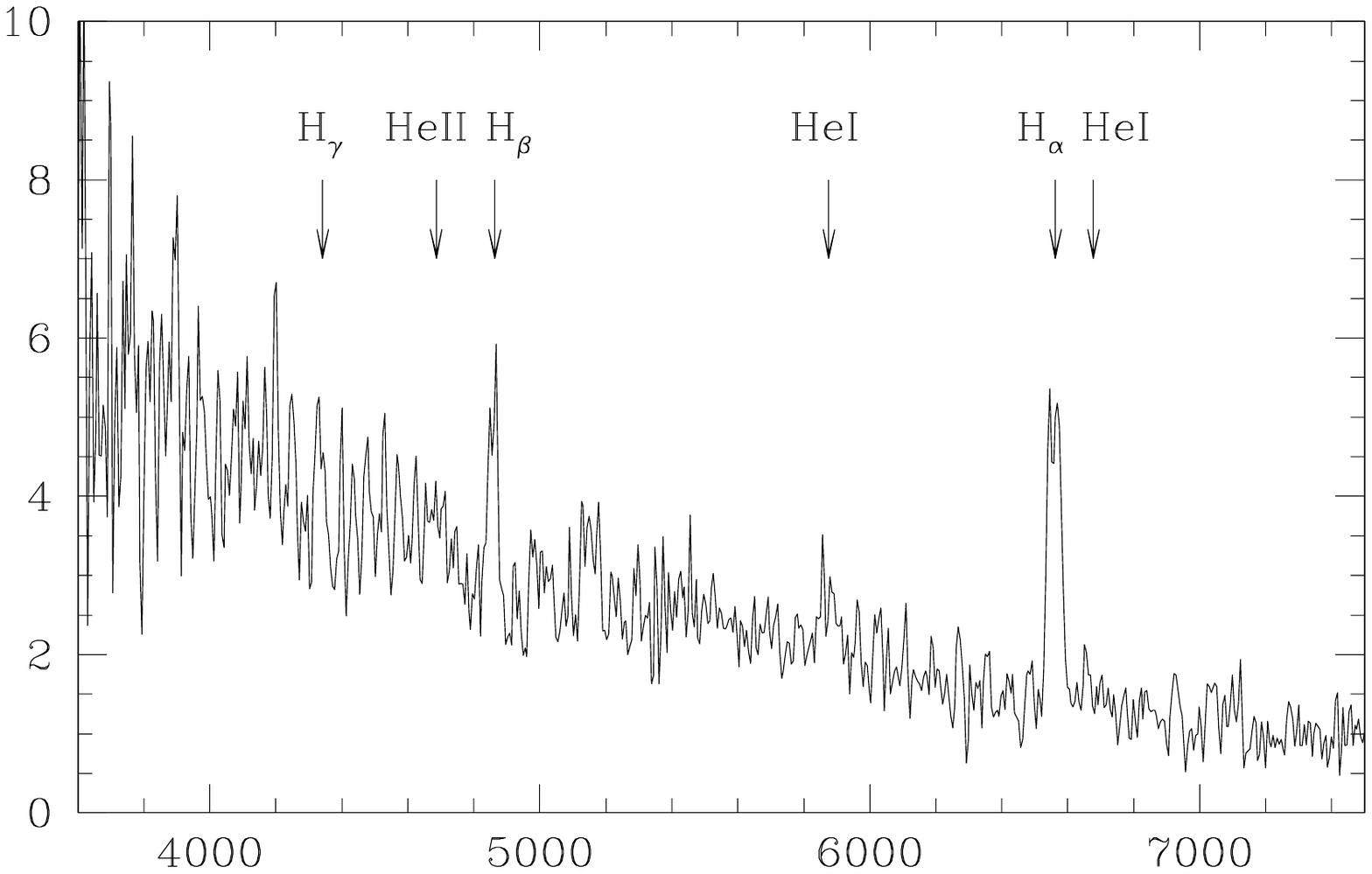}{$\lambda$, \AA}{$F_\lambda, \times
      10^{-17}$~erg~s$^{-1}$~cm$^{-2}$~\AA$^{-1}$} \\
    
  \end{tabular}
  \caption{--- Spectra of CVs detected in the 400d survey.}
  \label{fig:spectra}
\end{figure*}

\section{Optical observations of CV candidates}

The additional optical observations were carried out with the
Russian--Turkish 1.5-m telescope (RTT-150) using the medium- and
low-resolution TFOSC spectrograph and the 6-m (BTA) telescope at the
Special Astrophysical Observatory of the Russian Academy of Sciences,
where the SCORPIO-2 spectrograph \citep{scorpio05,scorpio11} was used
for the observations. Recently, beginning in the fall of 2015, the
observations have also been carried out with the 1.6-m AZT-33IK
telescope at the Sayan Observatory of the Institute of
Solar-Terrestrial Physics, the Siberian Branch of the Russian Academy
of Sciences, using the new medium- and low-resolution ADAM
spectrograph \citep{adam16,adamsrg16}. In all cases, we used grisms or
volume phase holographic gratings optimized for a range 4000--6000\AA,
which also includes the blue part of the spectrum starting from
3500--3700\AA. The spectral resolution was from 7\AA\ for SCORPIO-2 at
the BTA telescope to 12\AA\ for TFOSC at the RTT-150.

The spectroscopic observations of objects from this programme were
carried out during the last few years. We observed 38 objects, four of
which turned out to be new, previously unknown CVs. The spectra for
two of them (400d~j$001912.9\!+\!220736$ and
400d~j$152212.8\!+\!080338$) were later also obtained in the SDSS. The
remaining objects turned out to be quasars at various redshifts,
typically, at $z=1$--$2$. Examples of the spectra for such quasars are
given in \cite{ayut15}. The spectra of CV found in our work are
presented in Fig.~\ref{fig:spectra}. The spectra of 15 objects with
magnitudes $19.5<g^\prime<20$ were not measured to date; this work
will be continued.

\begin{table*}
  \begin{center}

    \caption{Cataclysmic variables detected in 400d survey} 
    \label{tab:res}
    \renewcommand{\arraystretch}{1.1}
    \renewcommand{\tabcolsep}{0.2cm}
    \footnotesize
    \begin{tabular}{lcccccl}
      \noalign{\vskip 3pt\hrule \vskip 1.5pt \hrule \vskip 5pt}
      Name & $\alpha$ (J2000) & $\delta$  (J2000) & $m_{g^\prime}$ & $f_X$, ~erg\,s$^{-1}$cm$^{-2}$ & $L_X$, ~erg\,s$^{-1}$ & Other name\\
      \noalign{\vskip 3pt\hrule\vskip 3pt}
      
      400d~j$001912.9\!+\!220736$ & $00~19~12.6$ & $+22~07~33$ & $19.61$ & $4.84\cdot 10^{-14}$ & $^*6.4\cdot 10^{29}$ & SDSS J001912.58+220733.0 \\ 
400d~j$050146.2\!-\!035914$ & $05~01~46.4$ & $-03~59~20$ & $18.41$ & $1.78\cdot 10^{-13}$ & $5.8\cdot 10^{30}$ & HY Eri                             \\ 
400d~j$124325.7\!+\!025541$ & $12~43~25.9$ & $+02~55~47$ & $17.72$ & $5.51\cdot 10^{-13}$ & $^*1.3\cdot 10^{30}$ & 1E 1240.8+0312 \\ 
400d~j$152212.8\!+\!080338$ & $15~22~12.2$ & $+08~03~41$ & $18.99$ & $1.58\cdot 10^{-13}$ & $^*1.2\cdot 10^{30}$ & SDSS J152212.20+080340.9  \\ 
400d~j$154730.1\!+\!071151$ & $15~47~30.8$ & $+07~11~31$ & $16.34$ & $1.16\cdot 10^{-13}$ & $2.0\cdot 10^{29}$ &  \\ 
400d~j$160002.4\!+\!331120$ & $16~00~03.7$ & $+33~11~14$ & $19.89$ & $8.87\cdot 10^{-14}$ & $^*1.5\cdot 10^{30}$ & VW CrB \\ 
400d~j$160547.5\!+\!240524$ & $16~05~48.0$ & $+24~05~31$ & $19.78$ & $5.47\cdot 10^{-14}$ & $^*8.5\cdot 10^{29}$ &  \\ 
400d~j$204720.3\!+\!000008$ & $20~47~20.8$ & $+00~00~08$ & $19.36$ & $4.19\cdot 10^{-14}$ & $^*4.4\cdot 10^{29}$ & SDSS J204720.76+000007.7  \\

      \noalign{\vskip 2pt\hrule\vskip 2pt}
    \end{tabular}
  \end{center}
  \footnotesize
  
  \vskip -8pt
  $^*$ --- The rough estimate of the X-ray luminosity $L_X$ that we use for our preliminary
  estimates of the luminosity function (see text for more details)
  
\end{table*}

\begin{figure}
  \centering
  \vskip -2mm
  \includegraphics[width=0.95\linewidth,viewport=31 168 594
  712]{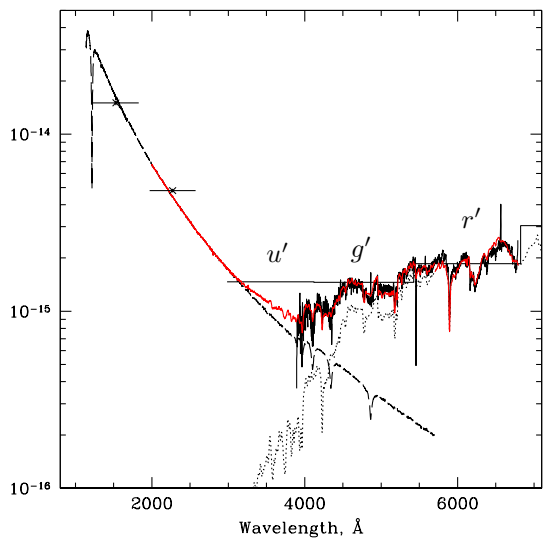}
  \vskip -4.3cm \hskip 0cm $u^\prime$
  \vskip -0.5cm \hskip 2.2cm $g^\prime$
  \vskip -0.8cm \hskip 5.1cm $r^\prime$
  \vskip 4.2cm
  \caption{Broadband spectrum of the CV
    400d~j$154730.1\!+\!071151$. The black curve shows the RTT-150
    spectrum (see Fig.~\ref{fig:spectra}). The red curve shows the
    model spectrum of a binary system which consist of an M2\,V red
    dwarf and a white dwarf with a temperature of
    $\approx 3\cdot 10^4$~K (WD0320$-$539). The measurements in the
    range 1500--2500\AA\ are taken from the observations with the
    GALEX orbital ultraviolet telescope \citep{martin05}. The
    histogram also indicates the photometric SDSS measurements in the
    $u^\prime g^\prime r^\prime i^\prime$ bands.}
  \label{fig:spec_j1547}
\end{figure}

\begin{figure}
  \centering \smfigure{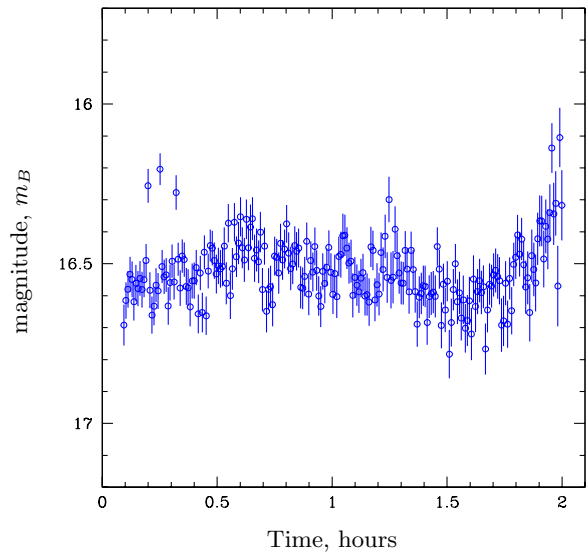}{Time, hours}{magnitude,
    $m_B$}
  \caption{B-band light curve of the CV 400d~j$154730.1\!+\!071151$
    obtained with the Sayan observatory AZT-33IK telescope in March
    2013.}
  \label{fig:lc_j1547}
\end{figure}

\section{The CV Sample}

Basic information about the CVs in our sample is presented in the
Table~\ref{tab:res}. Four of them turn out to be previously known CVs,
we detected four more objects using the optical observations described
above, two of them were also independently detected in the SDSS during
the spectroscopic observations of quasar candidates
\citep{sdss12}. The spectra of the CVs detected in 400d survey are
presented in Fig.~\ref{fig:spectra}. The optical spectra of
400d~j$050146.2\!-\!035914$, 400d~j$124325.7\!+\!025541$,
400d~j$160002.4\!+\!331120$ и 400d~j$204720.3\!+\!000008$ can be found
in \cite{burwitz99}, \cite{stocke83}, \cite{liu99} and
\cite{szkody04}, respectively (for more details, see also below).

\subsection{Notes on individual objects}

\paragraph{400d~j$050146.2\!-\!035914$ (HY Eri)} 

The CV \emph{HY Eri}=\emph{RX J0501.7-0359} was discovered in the
ROSAT all-sky survey \citep{beuermann99}. It was found to be an
eclipsing binary system with a magnetized WD, a polar
\citep{burwitz99}. The overall spectrum of this system and its
measured orbital period, 2.855~h, are also given in \cite{burwitz99}.

Due to the absence of an accretion disk in the binary system, its
optical brightness in the blue part of the spectrum is determined
mainly by the emission from its hot WD. By assuming certain
temperature ($10^4$~K) and mass ($0.6$--$0.8~M_\odot$) of the WD and,
consequently, its absolute magnitude $M_{g^\prime}\approx 11$--$12$,
we can estimate the distance to the binary system,
$D \approx 200$--$300$~pc. 

The binary system was detected in the infrared in the 2MASS all-sky
survey with $J=16.67$ \citep{skrutskie06}. In this part of the
spectrum, the companion star makes a major contribution.  Using the
absolute magnitudes of CV companion stars tabulated in
\cite{knigge06}, we can make another distance estimate for the binary
system: $D\approx 520$~pc. In fact, the first distance estimate is
only a lower limit, because the possible strong heating of the WD
surface through accretion is not taken in account; therefore, the
second distance estimate should be used.

\paragraph{400d~j$124325.7\!+\!025541$ (1E 1240.8+0312)} 

The source 1E 1240.8+0312 was discovered in the Einstein Observatory
Medium Sensitivity Survey and identified with an optical object with
bright emission lines \citep{stocke83}. In the same paper, it was
suggested that the source is a dwarf nova, i.e., a CV with a WD
without a strong magnetic field.

\paragraph{400d~j$154730.1\!+\!071151$}

A broadband spectrum of another CV from our survey,
400d~j$154730.1\!+\!071151$, is presented in
Fig.~\ref{fig:spec_j1547}.  We approximately decomposed the optical
RTT-150 spectrum into two components corresponding to an M2\,V red
dwarf, whose spectrum was taken from the stellar spectral flux library
of \cite{pickles98}, and a white dwarf (the spectrum WD0320-539 from
the HST CALSPEC
library\footnote{http://www.stsci.edu/hst/observatory/crds/calspec.html}). This
model also describes well the ultraviolet measurements with the GALEX
orbital ultraviolet telescope \citep{martin05}.

In Fig.~\ref{fig:lc_j1547} we show the B-band light curve of this CV
obtained with Sayan Observatory AZT-33IK telescope in March 2013. The
light curve of this object exhibits variability at a level of about
30\% of the flux. Therefore, apart from the white and red dwarfs, an
appreciable contribution of the emission from the accretion flow
around the WD must also be present in the emission from this object
near the B photometric band.

Unfortunately, in SDSS this object turns out to be in saturation in
$r^\prime, i^\prime$ and $z^\prime$ bands. However, in
Fig.~\ref{fig:spec_j1547} one can see that the $r^\prime$ magnitude
is, nevertheless, approximately correct, and it can be used for rough
estimates. Since the contributions of the WD and the accretion disk to
the total flux from the binary system are insignificant in the
$r^\prime$ band and assuming that the red dwarf in our binary system
is a main-sequence star, we can roughly estimate the distance.  The
absolute magnitude of M2\,V red dwarf in $r^\prime$ band is
$M_{r^\prime}\approx 10.1$, the source observed brightness is
$r^\prime=15.47$, therefore, the distance to the source can be
estimated to be about 120~pc.

Note, that the X-ray luminosity of the binary system based on this
distance estimate is $L_X\approx 2 \cdot 10^{29}$~erg~s$^{-1}$, which
roughly corresponds to the accretion rate of $5\cdot10^{-14}~M_\odot$
for a WD of mass $\approx 0.5~M_\odot$ (if the X-ray luminosity
accounts for a significant fraction of the bolometric luminosity of
the accretion flow). Such an accretion rate is too low for a
Roche-lobe-filling M2\,V companion star. Thus, this object may be
classified as the so-called ``pre-cataclysmic variable'', i.e., a
binary system where the mass transfer from the companion star onto the
compact star through the inner Lagrangian point has not yet completely
started.

\paragraph{400d~j$160002.4\!+\!331120$ (VW CrB)}

The CV VW\,CrB=Var21\,CrB was discovered at photographic plates of the
40-cm Sternberg Astronomical Institute astrograph at the Crimean
Astrophysical Observatory \citep{antipin96} as a dwarf nova that
exhibits the outbursts with nearly 3 mag amplitude,
$m_{\mathrm{phot}} = 14.5$--$17.5$, 10--15 days in duration.
Subsequently, the orbital period of a binary system was measured to be
107 min \citep{novak97,rk03}. The spectrum of this system, obtained
during the outburst, at the object's magnitude $m_B = 15.8$, can be
found in \cite{liu99}.

\paragraph{400d~j$204720.3\!+\!000008$} 

This object was discovered in SDSS \citep{szkody04}. This CV was
identified as a dwaft nova with 4.475~h orbital period \citep{rk03}.

\section{CONSTRAINTS ON THE CV X-ray LUMINOSITY FUNCTION IN THE SOLAR
  NEIGHBORHOOD}

As it was discussed above, we detected eight CVs, with more or less
reliable distance estimates being available only for two of them. To
obtain rough, preliminary estimates of the constraints on the
luminosity function, let us assume that the remaining systems have a
low X-ray luminosity. In this case, their absolute magnitudes must be
near minimum value of $M_{g^\prime}\approx12$ which is defined by the
presence of a WD in the system \citep{mikej14}. In these systems, the
accretion rate onto the WD must be low, and their orbital periods must
be not far from the period minimum. Therefore, this value of
$M_{g^\prime}$ agrees well with the estimate of
$M_{g^\prime}\approx11.6$ for systems with periods near the minimum
\citep{gansicke09}.

Note, that signatures of WD emission are clearly seen in the spectra
of two systems (400d~j$001912.9\!+\!220736$ и
400d~j$160547.5\!+\!240524$, see Fig.~\ref{fig:spectra}), which is
expected for such systems \citep{gansicke09}. The system
400d~j$160002.4\!+\!331120$ is known to have a short orbital period,
$P_{\mathrm{orb}} = 107$~min, which is quite close to the period
minimum (86 min). Therefore, its absolute magnitude should also be
close to the above one.

The X-ray luminosities that are derived for systems with unknown
distances under the assumption of $M_{g^\prime}\approx12$ are given in
the table and marked by asterisks. The distance estimates for these
systems turn out to be within the range 200--300~pc. Some of the
systems (400d~j$124325.7\!+\!025541$, 400d~j$152212.8\!+\!080338$ and
400d~j$204720.3\!+\!000008$ ) can actually have a higher
luminosity. However, the luminosity of all these systems, on average,
cannot be much higher, because, in this case, the objects should be
located at appreciably larger distances, but at high Galactic
latitudes the space density of sources at distances larger than the
thickness of the Galactic disk drops rapidly.

\begin{figure}
  \centering \smfigure{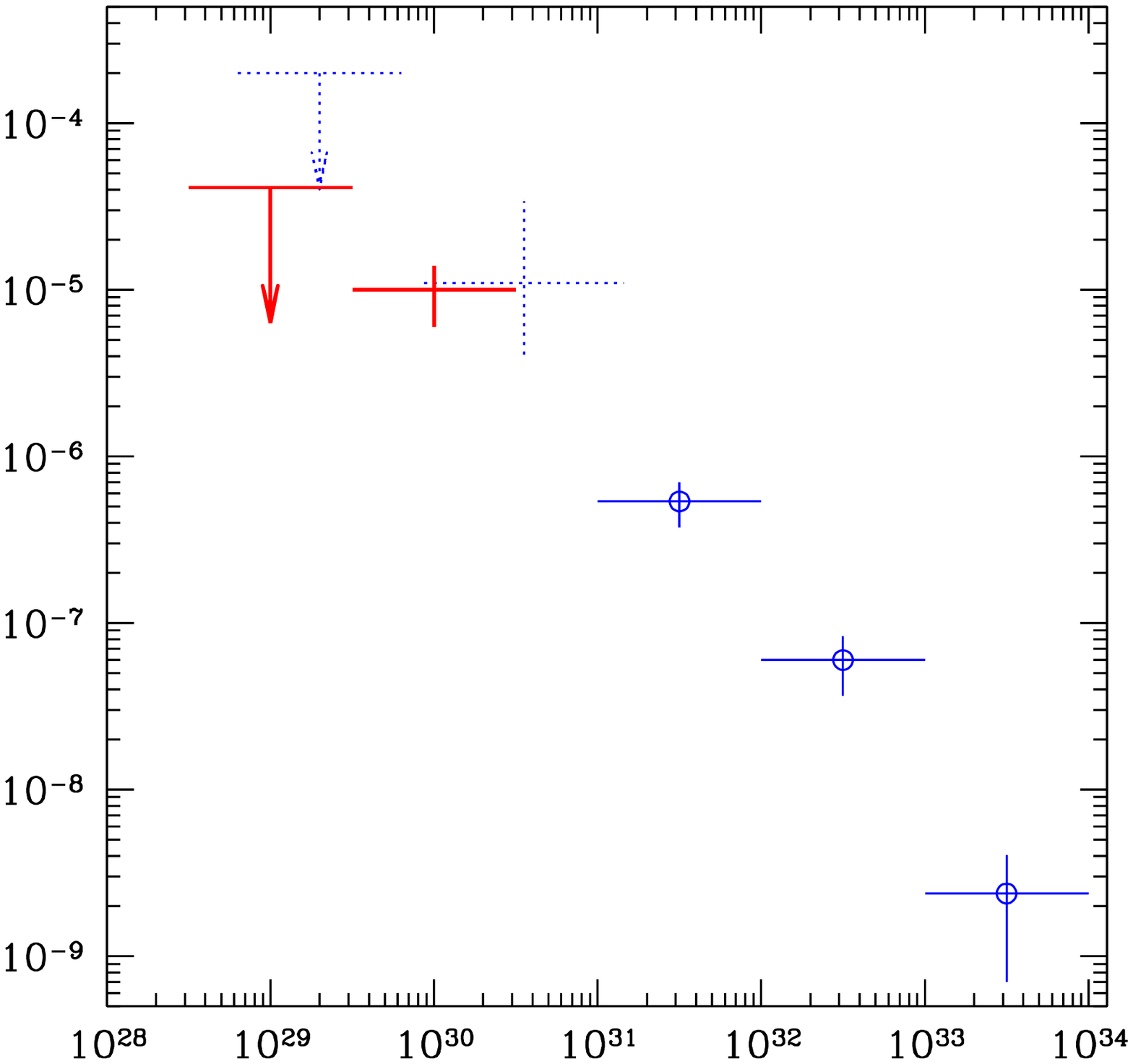}{$L_X$
    ($0.5$-$2$~keV),~erg\,s$^{-1}$}{$dN/d \lg L_X$, pc$^{-3}$}
  \caption{Constraints on the CV X-ray luminosity function. Our
    constraints are shown with the red crosses. In addition, the blue
    circles show the measurements of the X-ray luminosity function
    based on RXTE data \citep{sazonov06} recalculated to the
    0.5--2~keV energy band under the assumption of a power-law spectrum
    $dN/dE\propto E^{-1.6}$. The dotted blue crosses indicate the
    constraints on the luminosity function from the the ROSAT North
    Ecliptic Pole Survey \citep{pretorius07b}.}
  \label{fig:lf}
\end{figure}

Based on these data, we can obtain preliminary constraints on the CV
X-ray luminosity function. For simplicity, we will assume that the
flux limit for X-ray sources in the 400d survey is
$2.5\cdot10^{-14}$~erg\,s$^{-1}$cm$^{-2}$ irrespective of the ROSAT
pointing field, and that the area overlapping with the SDSS is 262.3
sq. deg. The dependence of the density of sources in the Galactic disk
on the perpendicular distance from the Galactic plane,
$\rho (z) = \rho_0 e^{-|z|/h}$, can be taken into account trough the
calculation of generalized volume, as it was done by \cite{tinney93}
and \cite{pretorius07b}. We will assume that the exponential scale
height of the Galactic disk for our CVs is 260~pc, which corresponds
to the disk scale height for old, short-period systems
\citep{pretorius07a}.

It should be also noted that only systems with $m_{g^\prime}<20$ are
included in our sample. This constraint does not affect the selection
of systems located at distances less than about 400~pc, because there
is a minimum absolute magnitude $M_{g^\prime}\approx12$ for CVs
\citep{mikej14}. For systems located at larger distances, the
selection probability depends not only on the X-ray flux but also on
the $g^\prime$ magnitude.  Therefore, the system
400d~j$050146.2\!-\!035914$ (HY Eri), whose distance can be estimated
to be 520 pc, is not taken in account in the estimates below.

The sample incompleteness can be taken into account statistically,
appropriately increasing the measurement errors. As was pointed out
above, at present we carried out optical observations for 21 out of
the 36 selected objects with magnitudes $19.5<g^\prime<20$, with two
previously unknown CVs having been detected among them. Therefore, the
number of previously unknown CVs among the remaining 15 objects can be
estimated to be, on average, about 1.4. It can be seen from the table
that these CVs will most likely have luminosities
$L_X\approx 10^{30}$~erg\,s$^{-1}$.

Six systems from our sample are found in in the range of X-ray
luminosities $3\cdot 10^{29}$--$3\cdot10^{30}$~erg\,s$^{-1}$. Taking
into account all mentioned above, we obtain the following constraint
on the CV space density based on this sample:
$\rho_0 = 1.0\pm 0.4 \cdot 10^{-5}$~pc$^{-3}$. Only one system,
400d~j$154730.1\!+\!071151$, with a luminosity close to the upper
boundary of the X-ray luminosity range
$3\cdot 10^{28}$--$3\cdot10^{29}$~erg\,s$^{-1}$ is found in this
luminosity range. In this case, the upper limit on the CV space
density for this range is $\rho_0 < 4.1 \cdot 10^{-5}$~pc$^{-3}$ (at
95\% confidence). However, this system is apparently a pre-cataclysmic
variable. If such systems are not considered as CVs, then no object is
found in the luminosity range
$3\cdot 10^{28}$--$3\cdot10^{29}$~erg\,s$^{-1}$ and the limit on the
CV space density is then $\rho_0 < 2.7 \cdot 10^{-5}$~pc$^{-3}$.

Our constraints on the X-ray luminosity function of CVs are shown with
solid red crosses in Fig.~\ref{fig:lf}; the blue circles indicate the
measurements of the X-ray luminosity function based on RXTE data
\citep{sazonov06} recalculated to the 0.5--2~keV energy band under the
assumption of a power-law spectrum $dN/dE\propto E^{-1.6}$; the dotted
crosses show the constraints on the luminosity function from the ROSAT
North Ecliptic Pole survey \cite{pretorius07b}. It can be seen from
the figure that our constraint on the CV space density near
$L_X\approx 10^{30}$~erg\,s$^{-1}$ agrees well with that from
\cite{pretorius07b}. Note that we obtain significantly stronger upper
limit on the CV space density near an X-ray luminosity
$L_X\approx 10^{29}$~erg\,s$^{-1}$.

\section{Discussion}

We present a sample of CVs detected among the X-ray sources of the 400
square degree (400d) survey. As it was shown above, this X-ray survey
is well suited to search for low X-ray luminosity CVs. Note that among
the systems selected in 400d survey, there is an appreciable fraction
of systems whose spectra clearly show signatures of the WD emission
(see Fig.~\ref{fig:spectra}), which is expected for systems with a low
accretion rate \citep{gansicke09}.

Based on this X-ray sample of CVs, we obtained preliminary constraints
on the X-ray luminosity function of CVs in the range of low X-ray
luminosities, $L_X\approx 10^{29}$--$10^{30}$~erg\,s$^{-1}$.  From the
comparison of our constraints with the luminosity function at
$L_X> 10^{31}$~erg\,s$^{-1}$ measured previously \citep{sazonov06}, we
conclude that its logarithmic slope is less steep at low luminosities
(see Fig.~\ref{fig:lf}).

The presented results are preliminary ones. They can be significantly
improved if more or less reliable distance estimates will be obtained
for a larger number of systems and if the statistically complete
sample will be expanded to include system with a lower optical
brightness. We are going to continue this work in future.

Our constraints on the CV luminosity function can be compared with
theoretical estimates of the space density of these objects. For
example, de Kool (1992) estimated the space density of all CVs to be
$\approx 0.5$--$2 \cdot 10^{-4}$~pc$^{-3}$ (recalculated to the Galaxy
age of 10 Gyr). As follows from Fig.~\ref{fig:lf}, we can eliminate
such a number of CVs in the investigated luminosity range. This may
imply, for example, that a large number of CVs have an even lower
X-ray luminosity. It should also be taken in account that the CV
luminosity can change in a wide range due to accretion instability.

From our estimates of the CV X-ray luminosity function, we conclude
that several thousand CVs will be detected in the future
Specrtum-R\"oentgen-Gamma all-sky X-ray survey with eROSITA telescope
at high Galactic latitudes, which will allow to significantly improve
the measurements of their X-ray luminosity function at X-ray
luminosities $L_X< 10^{31}$~erg\,s$^{-1}$. Further studies of CV
samples at lower luminosities, $L_X < 10^{29}$~erg\,s$^{-1}$, will
possibly allow to detect large number of systems at late post-bounce
evolutionary stages and to measure their space density.

\acknowledgements 

This work was supported by the Russian Foundation for Basic Research
(project no.\ 13-02-00741-a) and grant NSh-6137.2014.2. The
calculations of the area of the 400d X-ray survey used here were
supported by RSF grant no.\ 14-22-00271. The observations at the
6-meter BTA telescope were carried out with the financial support of
the Ministry of Education and Science of the Russian Federation
(agreement no. 14.619.21.0004, project ID RFMEFI61914X0004).

\end{document}